\documentclass[endnotes,12pt,aps]{article}
\usepackage[T1]{fontenc}
\usepackage{graphics}
\usepackage{amsmath}
\usepackage{amssymb}
\usepackage{graphicx,float}
\usepackage{epsfig,float}
\usepackage{latexsym}
\usepackage{epsf} 
\usepackage{graphicx}
\usepackage{epsfig}
\begin{document}
\renewcommand{\thefootnote}{\fnsymbol{footnote}}
\renewcommand{\thesection}{\Roman{section}}
\renewcommand{\thesubsubsection}{\arabic{subsubsection}.}
\def\figurename{\protect{\textbf{Figure}}}
\def\tablename{\protect{\textbf{Table}}}
\roman{section}
\date{}

\baselineskip=20pt

\title{\textbf{\textsf{\Large Equation of state, structure and diffusion \\ 
coefficients of Gay-Berne fluids: the cases \\ $\kappa'=5,10,15,20$}}}

\author{Jos\'e Guillermo M\'endez-Berm\'udez $^\dagger$ , \,
Iv\'an Guill\'en-Escamilla$^\dagger$, \\ Juan Carlos Mixteco-S\'anchez $^\dagger$, Gloria Arlette M\'endez-Maldonado $^\ddagger$
, \\ Minerva Gonz\'alez-Melchor$^\S$ \footnote{ Authors to whom 
correspondence should be addressed. E-mail:minerva@ifuap.buap.mx, jose.mendez@valles.udg.mx}\\
{\it{\small $^\dagger$ Centro Universitario de los Valles, Universidad de Guadalajara,}} \\
{\it{\small Carretera Guadalajara-Ameca Km. 45.5, Ameca 46600, Jalisco, M\'exico.}} \\ 
{\it{\small $^\ddagger$ Centro Universitario de Ciencias Exactas e Ingenier\'ia, Universidad de Guadalajara}} \\
{\it{\small Blvd. Marcelino Garc\'ia Barrag\'an No. 1412, Guadalajara 44430, Jalisco, M\'exico.}} \\
{\it{\small $\S$ Instituto de F\'isica "Luis Rivera Terrazas" }} \\
{\it{\small Benem\'erita Universidad Aut\'onoma de Puebla, Apdo. Postal J-48, 72570 Puebla, M\'exico.}}
}

\maketitle

\begin{abstract}
\baselineskip=18pt
{\small
We performed extensive molecular dynamics simulations to obtain pressure-density phase diagram, orientational order parameter, pair correlation functions and translational diffusion coefficients of Gay-Berne fluids. Different sets of parameters were employed for the Gay-Berne potential, in particular we studied the cases $\kappa'=5, 10, 15, 20$ and $\kappa=3$, $\mu=2$, $\nu=1$ at different conditions of density and temperature. The structure was analyzed in terms of the order parameter and the pair correlation functions. We found that for the highest value $\kappa'=20$ the region where pressure increases with density is significantly reduced at low temperatures; additionally the pressure shows several decays with density as an indicative that several structural phases can take place. These effects are discussed in terms of the pair correlation functions. For higher temperatures the pressure shows only two decays for all $\kappa'$'s studied. As this parameter increases its value those decays are shifted to lower densities.}\\



\end{abstract}

\vspace{5pt}

\section{Introduction}

Fluids composed of non-spherical molecules have been studied by using different methods and approaches. Liquid crystals, rod-like polymers, aqueous suspensions of tobacco mosaic virus (TMV) and disk-like particles have a high degree of shape anisotropy. This shape anisotropy allows them to exhibit a rich variety of structural phases as density and temperature are changed. In some thermotropic liquid crystals (for instance, the 4-n-pentylbenzenethio-4$'$-n-decyloxybenzoate) as temperature is increased the fluid undergoes different transitions going from the crystal to the smectics, nematic and isotropic phases \cite{book1}. Due to orientational and positional degrees of freedom these fluids exhibit phenomena not present in fluids of spherical particles \cite{rod-like}. They are use in many applications and with different purposes, going from the well known display technologies to medical devices in biological systems \cite{book1,app1} as the self-assembly of viruses in aqueous suspensions \cite{tmv1,tmv2,tmv3}.

Experiments, theoretical models and computer simulation studies have been conducted in recent years for pure fluids and mixtures of non-spherical particles \cite{mix1,mix2,mix3,mix4}. However these studies are scarce compared to the research that has been done in fluids of spherical particles, where phase diagrams, static structure and dynamic properties have received wide attention. By the theoretical side many studies have been undertaken on these systems \cite{teo1,teo2,teo3}, however is often more difficult to include orientational degrees of freedom and geometrical shape in theories than in numerical simulations \cite{guille}. Theoretical works have employed the Fokker-Planck equation \cite{guille,doi}, the generalized Langevin \cite{coffey,medina-01,medina-02}, the Onsager theory \cite{varga}, the density functional theory \cite{enrique4} and a generalized Van der Waals description \cite{VdWtheory}, among others. The non-spherical feature can be seen as an internal degree for instance a dipole orientation, whereas the geometrical shape often enters through some physical parameter as the diffusion coefficients without an explicit account of particle shape. 

Concerning computer molecular simulations, different techniques as Monte Carlo (MC) and molecular dynamics (MD) have allowed the calculations of phase behavior, thermodynamics, structure and dynamic properties of pure fluids of flexible, rigid and axially symmetric molecules \cite{allen-etal}. Fluids composed of spherocylinders (hard cylinders with semi-spherical endings) as another model for non-spherical particles are mainly studied by MC techniques \cite{carlos}, while models where the interaction potential is a continuous function of distance are suitable for MD calculations. From an atomistic point of view the intramolecular structure has been considered by including intramolecular sites interacting through bond, bending and torsional interactions. Other paths have employed the Yukawa and Lennard-Jones (LJ) potentials between sites \cite{tmv,sitesYuk2}. However the atomistic approach often increases the size of the system.

From the models employed in literature, the Gay-Berne (GB) potential have played a crucial role in the description of mesophases in fluids of non-spherical particles \cite{pgb}. The GB model is flexible enough to allow the description of long ellipsoids, passing through spheres and ending in discotic particles, using one site per particle. This potential depends on four parameters for a pure fluid, usually denoted as ($\kappa,\kappa',\mu,\nu$), which are closely related to the shape of particles and the strength interaction between them. In this sense the GB model constitutes a family of potentials. From all possible sets of parameters the most studied is the ($\kappa=3, \kappa'=5, \mu=2, \nu=1$) GB fluid for ellipsoids, whose phase diagram, second rank orientational order parameter and pair distribution functions are already known \cite{enrique1}. Other properties have also been studied for this case as: the velocity autocorrelation function \cite{enrique2}, bulk and shear viscosities \cite{viscos}, elastic constants \cite{elastic1}, entalphy and free energies \cite{enrique4}, the isotropic-nematic transitions \cite{trans1} and the liquid-vapor coexistence \cite{trans2}. The GB potential also has been used to obtain the viscosities and stress \cite{no-new1} and self-diffusion coefficient \cite{no-new2} the non-Newtonian regime of different liquid crystal models. Recently the computer simulation have been used to understand nematic-vapour interface of the GB model for prolate molecules with $\kappa=4$ and $6$, and for oblate molecules $\kappa=0.3$ and $0.5$, whit different values of $\kappa'$ for each $\kappa$ \cite{trans2}; that together with the elastic properties of the liquid crystal are determinant to understand the formation of nematic droplets \cite{Rull_2012, Vanzano_2012, Vanzano_2016}.

Other sets of parameters have been explored under very specific conditions, in this direction Mori et al. \cite{shear-flow} examined the effect of changing $\nu$=1.8, 2.0, 2.2 on the orientational order parameter and the viscosities under a shear flow. By setting the values $\mu$=1 and $\nu$=3 and Germano et al. obtained elastic constants \cite{elastic2}. Bates and Luckhurst \cite{BL-1, BL-2} explored the values $\kappa=4.4, \kappa'=20, \mu=1, \nu=1$ and calculated the diffusion coefficients in the smectic A phase and De Miguel et al. \cite{enrique3} obtained stable smectic phases for the same parameters, and the pair distribution functions, phase diagrams and orientational order parameters were also calculated. Satoh investigated the rotational viscosity coefficients \cite{satoh1} and studied the effect of external magnetic fields \cite{satoh2}. In the case of very long particles ($\kappa=15$) the isotropic-nematic region was explored \cite{k15-01, k15-02}. The variation of the parameter $\kappa$ was done to analyze the isotropic-nematic region \cite{k-var3} by calculating orientational correlation functions.

The study of phases in discotic fluids has been explored by constructing columnar states \cite{disc-04}, varying the parameter $\kappa$ \cite{disc-05,disc-06} or varying both the energy strength and the geometrical parameters $\kappa$ and $\kappa'$ respectively, and the phase diagrams were obtained \cite{dicotic-01}. Other cases of study remained to be explored. In this direction the route that many studies have addressed is to parametrize the Gay-Berne potential for a particular type of molecules and adjust the set of parameters to reproduce the geometry or the interaction between pairs \cite{golubkov}. Computer simulations, by using GB potential, have been used to study discotic liquid crystal with $\kappa=0.345$, $0.2$, $1.0$, and $2.0$ \cite{Cienega_2014}, by resulting promising materials for technological applications in films that increase the angle of view in liquid crystal displays \cite{Bushbyand_2011}.

The variation of the parameter $\kappa'$ for ellipsoids have been analyzed taking the values $\kappa'$=1,5,6.63 and 8.33 to study the liquid-vapor region where $\kappa$ was set to $\kappa$=3 \cite{kp-var1}. The studied temperatures were $T^*=$0.5, 0.6, 0.65, 0.7 and 0.8. Another study of varying $\kappa'$ is found in \cite{kp-var2} where the pressure and the order parameters were obtained for $\kappa'=$5, 10, 25 and $T^*=$0.7 and the authors analyzed the liquid-vapor region for $\kappa'=$5, 2.5, 1.25, 1. However a systematic study in terms of $(\kappa,\kappa',\mu,\nu)$ and different conditions of density and temperature to those already mentioned has not been performed, this is quite desirable to drawn the general phase behavior of this model.

In these work we have undertaken an extensive numerical study of non-spherical particle fluids by changing the interaction strength in the Gay-Berne model in a systematic way, covering regions where it has not been done. This allows us to quantify its effects on the pressure-density phase diagrams, the order parameter, the perpendicular and parallel correlation functions and the translational diffusion coefficients, both parallel and perpendicular to the director. This properties give us information of the smectic phase. The rest of the paper is organized as follows: Section II contains a brief description of the Gay-Berne interaction potential, section III summarizes details concerning the procedure followed in the simulations. In section IV we present the definition of the properties. The results on phase diagrams, order parameter, radial distribution functions and diffusion coefficients are presented and discussed in section V. Finally, conclusions are given in section VI.

\section{Gay-Berne potential model}

The Gay-Berne potential was introduced as a model to simulate the interaction between two elongated four-site Lennard-Jones molecules through an effective pair potential between two particles with no internal structure. This model was proposed by Gay and Berne \cite{pgb} as a modification to the earlier Berne-Pechukas potential \cite{pechukas}, since then, the GB model have played a crucial role. This serves as a benchmark that accounts reasonably well for the shape of non-spherical particles. In this model two particles interact according to

\begin{equation}
u_{ij}(\hat{n}_{i},\hat{n}_{j},{\bf r})=4\epsilon(\hat{n}_{i},
\hat{n}_{j},{\bf r})\left[  \left( \frac{\sigma_{o}}{r-\sigma+\sigma_{o}} 
\right)^{12} - \left( \frac{\sigma_{o}}{r-\sigma+\sigma_{o}} \right)^{6} \right],
\label{potencial}
\end{equation}

\noindent where $\hat{n}_i$ is the orientational axial vector of particle $i$, $r = |{\bf r}|=|{\bf r}_i-{\bf r}_j|$ is the separation between center of mass of particles $i$ and $j$. The length $\sigma$ and strength $\epsilon$ are functions of the orientational vectors $\hat{n}_{i},\hat{n}_{j}$ and the separation $r$. These are given by

\begin{eqnarray}
\sigma & = & \sigma_{\mathrm{o}}\left\{ 1 -
\frac{\chi}{2r^{2}} \left[ \frac{\left({\bf r}\cdot\hat{n}_{i}+
{\bf r}\cdot\hat{n}_{j} \right)^{2}}{1+\chi\left(
\hat{n}_{i}\cdot\hat{n}_{j} \right) } +
\frac{ \left({\bf r}\cdot\hat{n}_{i}-{\bf r}
\cdot\hat{n}_{j} \right)^{2}}{ 1-\chi\left(
\hat{n}_{i}\cdot\hat{n}_{j} \right) }
\right] \right\}^{-1/2}\label{sigma},\\
\epsilon(\hat{n}_{i},\hat{n}_{j},{\bf r})
& = & 
\epsilon_{\mathrm{o}}\epsilon^{\nu}(\hat{n}_{i},\hat{n}_{j})
\epsilon^{\prime\mu}(\hat{n}_{i},\hat{n}_{j},{\bf r}),\label{epsilon}
\end{eqnarray}

\noindent where $\sigma_{\mathrm{o}}$ and $\epsilon_{\mathrm{o}}$ have length and energy units, respectively, and are used to make real quantities dimensionless, for spheres they are reduced to the usual LJ parameters. Additionally,

\begin{eqnarray}
\epsilon(\hat{n}_{i},\hat{n}_{j}) & = & \left[ 1 - \chi^{2}
\left( \hat{n}_{i}\cdot\hat{n}_{j} \right)^{2}  \right]^{-1/2},
\label{epsilon1}\\
\epsilon^{\prime}(\hat{n}_{i},\hat{n}_{j},{\bf r}) & = & 1 -
\frac{\chi^{\prime}}{2r^{2}} \left[ \frac{\left({\bf r}\cdot
\hat{n}_{i}+{\bf r}\cdot\hat{n}_{j} \right)^{2}}{1+\chi^{\prime}
\left( \hat{n}_{i}\cdot\hat{n}_{j} \right) } +  \frac{
\left({\bf r}\cdot\hat{n}_{i}-{\bf r}\cdot\hat{n}_{j}
\right)^{2}}{ 1-\chi^{\prime}\left( \hat{n}_{i}\cdot\hat{n}_{j}
\right) }  \right], \label{epsilon2}
\end{eqnarray}

\noindent where $\chi$ and $\chi^{\prime}$ are defined as

\begin{equation}
\chi=\frac{\kappa^2-1}{\kappa^2+1}
\hspace{0.5cm}\mbox{and}\hspace{0.5cm} 
\chi^{\prime}=\frac{\kappa^{\prime(1/\mu)}-1}{\kappa^{\prime(1/\mu)}+1},
\end{equation}

\noindent $\chi$ is a function of the shape anisotropy parameter $\kappa$, which describes the particle shape (rodlike $\kappa > 1$, disc-like $\kappa < 1$, spheres $\kappa =1$) and $\chi^{\prime}$ is a function of the energy anisotropy parameter, $\kappa^{\prime}$, this letter provides the ratio between the potential well depths for the side-side and end-end configurations. In this way the Gay-Berne potential is usually specified in the form GB ($\kappa$,$\kappa'$,$\mu$,$\nu$). The exponents $\mu$ and $\nu$ usually take values 2 and 1, respectively \cite{pgb}, although other values have also been used. 

\vspace{1.0cm}
\begin{figure}[htb!]
\centerline{\epsfig{figure=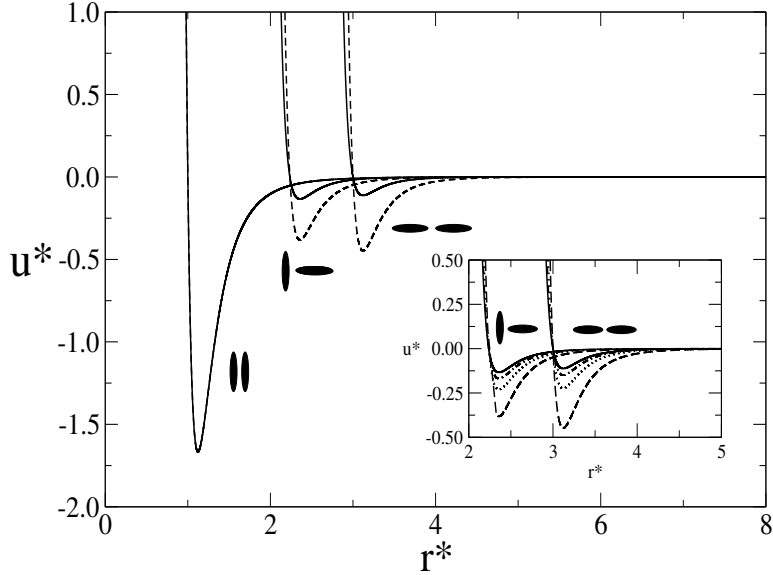,width=4.0in}}
\caption{The Gay-Berne potential for side-side, ``T'' and end-end configurations at $k'=5$ (dashed lines) and $k'=20$ (solid lines). The inset, the Gay-Berne potential for ``T'' and end-end configurations at $k'=5$ (dashed lines), $k'=10$ (dot lines), $k'=15$ (dot-dashed lines) and $k'=20$ (solid lines).} \label{pgb-01}
\end{figure}

In this work we study GB fluids for different values of $\kappa'$ and fixed $\kappa$ = 3, $\mu$ = 2, $\mu$ = 1, we denoted them by GB (3,$\kappa'$,2,1) in regions of temperature where it has not been done, in particular we set $\kappa'$ = 5, 10, 15, 20 at temperatures $T^*$ = 0.5, 0.75, 1.0, 1.25. The case $\kappa'$ = 5 was studied for comparison with previous work \cite{enrique1}. The GB potential is shown in Fig. \ref{pgb-01} for $\kappa'$ = 5, 20, the inset of the figure displays the ``T'' and the end-end configurations for $\kappa$'=5,10,15,20, the side-side configuration (not shown) has not change by varying the $\kappa$' parameter. We calculated pressure-density phase diagram, the order parameter, the parallel and perpendicular correlation functions and the translational diffusion coefficients for these four GB fluids.

\noindent In order to solve Newton's equations of motion in the molecular dynamics, the force between a pair of GB particles $i$ and $j$, is calculated according to 

\begin{equation}
{\bf F}_{ij}=-\frac{\partial u}{\partial r}{\bf r}_{ij}-
\frac{\partial u}{\partial a}\hat{n}_{i}-
\frac{\partial u}{\partial b}\hat{n}_{j},
\label{fuerza1}
\end{equation}

\noindent where we have defined  $a={\bf r}\cdot\hat{n}_{i}$, $b={\bf r}\cdot\hat{n}_{j}$, $c=\hat{n}_{i}\cdot\hat{n}_{j}$. The force ${\bf F}_{ij}$ obeys ${\bf F}_{ij}=-{\bf F}_{ji}$ and the torque has to be calculated as \cite{antypov}

\begin{eqnarray}
\boldsymbol{\tau}_{ij} & = & -\hat{n}_{i}\times\left(
\frac{\partial u}{\partial a}{\bf r}_{ij} +
\frac{\partial u}{\partial c}\hat{n}_{j} \right)\label{torca1},\\
\boldsymbol{\tau}_{ji} & = & -\hat{n}_{j}\times\left( \frac{\partial u}
{\partial b}{\bf r}_{ij} + \frac{\partial u}
{\partial c}\hat{n}_{i} \right)\label{torca2},
\end{eqnarray}

\noindent where the partial derivatives are given by

\begin{eqnarray}
\frac{\partial u}{\partial r} & = & 4 \epsilon(\hat{n}_{i},\hat{n}_{j},
{\bf r}) \left\{ \frac{A\mu\chi^{\prime}}{\epsilon^{\prime}
(\hat{n}_{i},\hat{n}_{j},{\bf r})r^{4}} \left[ \frac{(a+b)^{2}}
{1+\chi^{\prime}c} + \frac{(a-b)^{2}}{1-\chi^{\prime}c} \right]
\right.\nonumber\\
& & \left. - \frac{\sigma^{3}\chi B}{2r^{4}} \left[
\frac{(a+b)^{2}}{1+\chi c} + \frac{(a-b)^{2}}{1-\chi c}
\right] - \frac{B}{r}  \right\},
\end{eqnarray}

\begin{eqnarray}
\frac{\partial u}{\partial a} & = & 4 \epsilon(\hat{n}_{i},
\hat{n}_{j},{\bf r}) \left\{ -\frac{A\mu\chi^{\prime}}
{\epsilon^{\prime}(\hat{n}_{i},\hat{n}_{j},{\bf r})r^{2}}
\left[ \frac{a+b}{1+\chi^{\prime}c} +
\frac{a-b}{1-\chi^{\prime}c} \right]\right.\nonumber\\
& & \left. + \frac{\sigma^{3}\chi B}{2r^{2}}
\left[ \frac{a+b}{1+\chi c} + \frac{a-b}{1-\chi c} \right] \right\},
\end{eqnarray}

\begin{eqnarray}
\frac{\partial u}{\partial b} & =
& 4 \epsilon(\hat{n}_{i},\hat{n}_{j},{\bf r})
\left\{ -\frac{A\mu\chi^{\prime}}{\epsilon^{\prime}
(\hat{n}_{i},\hat{n}_{j},{\bf r})r^{2}} \left[
\frac{a+b}{1+\chi^{\prime}c} - \frac{a-b}{1-\chi^{\prime}c}
\right]\right.\nonumber\\
& & \left. + \frac{\sigma^{3}\chi B}{2r^{2}}
\left[ \frac{a+b}{1+\chi c} - \frac{a-b}{1-\chi c}
\right] \right\},
\end{eqnarray}

\begin{eqnarray}
\frac{\partial u}{\partial c} & = &
4 \epsilon(\hat{n}_{i},\hat{n}_{j},{\bf r})
\left\{ \frac{A\mu\chi^{\prime 2}}{\epsilon^{\prime}(\hat{n}_{i},
\hat{n}_{j},{\bf r})2r^{2}} \left[ \left(\frac{a+b}{1+
\chi^{\prime}c}\right)^{2} - \left(\frac{a-b}{1-\chi^{\prime}c}
\right)^{2} \right]\right.\nonumber\\
& & \left. + \frac{\sigma^{3}\chi^{2} B}{4r^{2}}
\left[ \left(\frac{a+b}{1+\chi c}\right)^{2}-
\left(\frac{a-b}{1-\chi c}\right)^{2} \right] \right.\nonumber\\
& & \left. + A\nu\chi^{2}(\hat{n}_{i}\cdot\hat{n}_{j})
\epsilon^{2}(\hat{n}_{i},\hat{n}_{j})\right\},
\end{eqnarray}

\noindent and the quantities $A$ and $B$ are defined as

\begin{eqnarray}
A&=&\left(\frac{\sigma_{o}}{r-\sigma+\sigma_{o}}\right)^{12}-
\left(\frac{\sigma_{o}}{r-\sigma+\sigma_{o}}\right)^{6},\nonumber\\
B&=&12\left(\frac{\sigma_{o}}{r-\sigma+\sigma_{o}}\right)^{13}-
6\left(\frac{\sigma_{o}}{r-\sigma+\sigma_{o}}\right)^{7}.
\end{eqnarray}

Given the force between a pair of particles, the equations of motion, both translational and orientational are solved to perform Molecular Dynamics of Gay-Berne fluids.

\section{Computer simulations}

We developed a MD simulation program at constant volume, number of particles and temperature \cite{allen}. The units of mass, length, and energy were chosen as $m$, $\sigma_{\mathrm{o}}$, and $\epsilon_{\mathrm{o}}$, respectively. We allocated $N=500$ particles in a cubic simulation box of volume $V^*=L^3$ $(V^*=V/\sigma_0^3)$. All particles were assigned inertial moment $I^*$=1 ($I^*=I(m\sigma_0^2)^{-1}$). Periodic boundary conditions and the minimum image convention were also employed, the cut-off distance was set to $r_c^*=L^*/2$ $(r_c^*=r_{c}/\sigma_0)$ in all cases. The temperature was kept constant by rescaling the velocities after each time step. The integration of the orientational and translational equations of motion was performed by using the Leap-Frog algorithm developed by Hockney and Potter \cite{hockney,potter} for the translational equations and by Fincham \cite{fincham} for the orientational motion. A time step of $\Delta t^*$ = 0.0015 $(\delta t^*=\Delta t(m\sigma_0^2/\epsilon_0)^{-1/2})$ was used to integrate the equations of motion. The initial configuration for each isotherm was prepared with particles fixed in a fcc lattice at a low-density $\rho^*$ = 0.005 $(\rho^*=\rho\sigma^3_0)$. Their random initial velocities obeyed the Maxwell-Boltzmann distribution \cite{max-boltz}. The unitary orientations and their derivatives were assigned randomly and obeyed a Gaussian distribution \cite{enrique1}. We used $1\times10^{4}$ time steps for the equilibration period and additional $2\times10^{4}$ iterations for calculating average properties. The data for each isotherm were generated starting from a low density state with $\rho^*=0.005$ from which the system was simulated, once the equilibrium was reached a run for production was conducted and the average properties were measured. With the final configuration the system was then compressed to obtain a new state of higher density. This procedure was repeated to obtain a full isotherm.

\vspace{1.0cm}
\begin{figure}[htb]
\centerline{\epsfig{figure=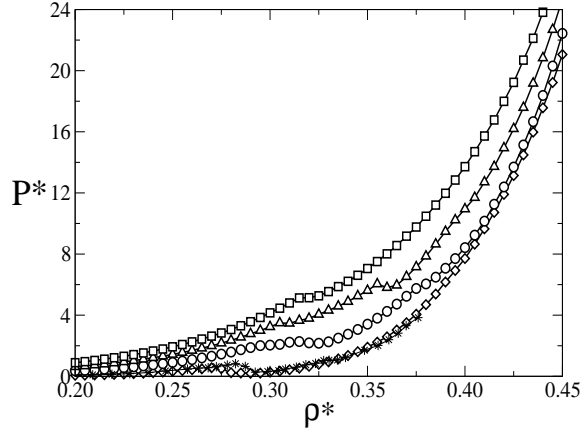,width=3.0in}}
\caption{Pressure-density phase diagrams for the (3,5,2,1) GB fluid for temperatures $T^*$ = 0.50 (diamonds), 0.75 (circles), 1.00 (triangles) and 1.25 (squares). The comparison with results from \cite{enrique1} for $T^*=0.5$ are shown with stars.}
\label{phase-3521}
\end{figure}

\section{Calculated properties}

\subsection{Pressure}

The pressure was calculated according to the virial expression as the sum of two contributions,

\begin{equation}
\langle P \rangle= \langle P^{kin} \rangle + \langle P^{int} \rangle,
\end{equation}

\noindent where $ P^{kin}$ and $P^{int}$ are
the kinetic and that due to forces between particles, given by

\begin{equation}
P^{kin} = \frac{1}{3V}\sum^N_{i=1}\left(m_i{\bf v}^2_i + I_i\boldsymbol{\omega}^2_i\right),
\end{equation}

\noindent where ${\bf v}_i$ and $\boldsymbol{\omega}_{i}$ are the translational and angular velocities of particle $i$, and

\begin{equation}
P^{int} \, = \frac{1}{V}\sum_{i=1} ^{N-1} \, \sum_{j>i} ^{N}{\bf r}_{ij}\cdot{\bf F}_{ij}~.
\label{virial}
\end{equation}

\subsection{Orientational order parameter}

The second rank orientational order parameter $P_2(t)$ gives the particle bulk orientational order, it takes values between 0 and 1. When $P_2=1$ the molecules are arranged in a crystal structure, whereas for $P_2=0$ the system is in an isotropic phase. The definition of $P_2(t)$ is given by

\begin{equation}
\langle P_2(t) \rangle=\left< \frac{1}{N}\sum^N_i P_2(\hat{n}_i(t)\cdot\hat{n}_d(t)) \right>,
\end{equation}

\noindent where $P_2$ is the second Legendre polynomial, the vector $\hat{n}_d$ is the director of the phase and $\left< ... \right>$ denote time averages. To obtain the director and the order parameter we maximize $P_{2}$ respect to all the rotations of $\hat{n}_d$ by writing ${\bf P}_{2}=\hat{n}_d\cdot{\mathbb Q}\cdot\hat{n}_d$, where ${\mathbb Q}$ is the ordering matrix. Thus we diagonalize the matrix ${\mathbb Q}$ that represents the orientational tensor, this matrix is defined by the $\alpha\beta$ element

\begin{equation}
Q_{\alpha\beta}=\frac{1}{2N}\sum^N_in_{i\alpha}n_{i\beta}-\delta_{\alpha\beta},
\end{equation}

\noindent where $n_{i\alpha}$ is the $\alpha-$component ($\alpha=x,y,z$) of $\hat{n}_i$ and $\delta_{\alpha\beta}$ is the Kronecker delta. The largest eigenvalue obtained by 
diagonalizing $Q_{\alpha\beta}$ is the order parameter and its corresponding eigenvector is defined as the director.

\begin{figure}[htp]
\centering \mbox{\includegraphics[width=4in]{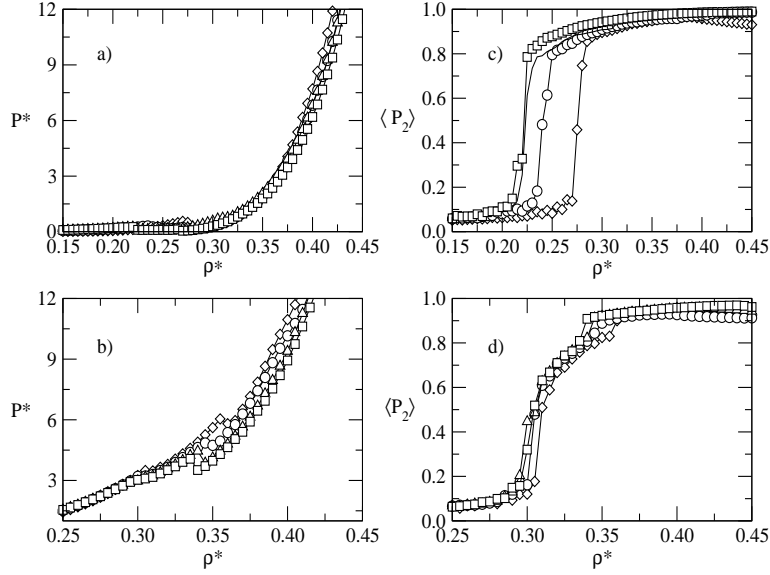}}
\caption{Pressure-density diagrams at two fixed temperatures: a) $T^*=$ 0.50 and b) $T^*=$ 1.00 and order parameter as function of density for the same temperatures: c) $T^*=$ 0.50 and d) $T^*=$ 1.00 for $\kappa'=$ 5 (diamonds), 10 (circles), 15 (triangles) and 20 (squares).}
\label{ppo-5101520}
\end{figure}

\begin{figure}[htp]
\centering \mbox{\includegraphics[width=4in]{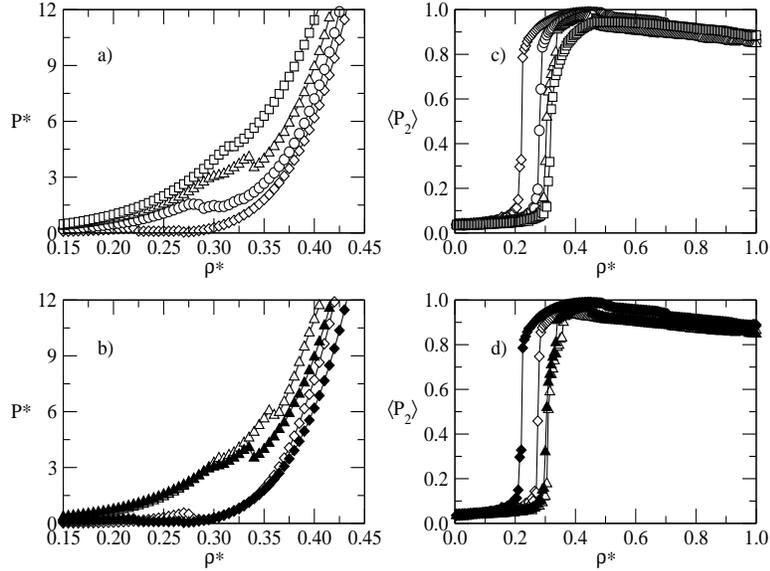}}
\caption{ a) Pressure-density diagram and c) Order parameter for the (3,20,2,1) GB fluid at temperatures $T^*$ = 0.50 (diamonds), 0.75 (circles), 1.00 (triangles) and 1.25 (squares). b) Pressure and d) Order parameter as functions of density at fixed temperatures $T^*$ = 0.50 (diamonds) and $T^*$ = 1.00 (triangles) for the (3,5,2,1) GB fluid (open symbols) and the (3,20,2,1) GB fluid (full symbols).}
\label{ppo-32021}
\end{figure}

\subsection{Pair correlation functions}

Besides the order parameter, a quantity useful in the classification of structural phases in fluids of non-spherical particles is the pair correlation function, $g(r)$, \cite{allen} which quantifies positional correlations. This function, $g(r)$, can be split into parallel and perpendicular contributions which are measured along the parallel and perpendicular components of the director, denoted by $r^*_{\|}$ and $r^*_{\bot}$, respectively. The parallel correlation function, $g(r^*_{\|})$, is useful for identifying smectic phases; because the layer structure of smectic phases shows up as a periodic variation of $g(r^*_{\|})$ while $g(r^*_{\bot})$ identifies smectic phases with in-layer order. Both are calculated for different states of the GB fluids.

\subsection{Translational diffusion coefficients}

Another quantity of interest in this work is the translational diffusion coefficients, two different relations can calculate this property: by using the mean square displacement (MSD) and via the velocity autocorrelation function (VACF) \cite{allen,frenkel}. We use the VACF to calculate the total, the parallel and the perpendicular translational diffusion coefficients with respect to the director. The total diffusion coefficient, defined as

\begin{equation}
D_{tr}=\frac{1}{3}\int^{\infty}_0 dt \left\langle {\bf v}_i(t)\cdot {\bf v}_i(0) \right\rangle,
\label{Dtr}
\end{equation}

\noindent can be splitted into parallel and perpendicular contributions, which in turn, define the parallel and perpendicular diffusion coefficients given by 

\begin{eqnarray}
D_{tr}^{\|}&=&\int^{\infty}_0 dt \left\langle {\bf v}^{\|}_i(t)\cdot {\bf v}^{\|}_i(0) 
\right\rangle,\label{Dtr(par)}\\
D_{tr}^{\bot}&=&\frac{1}{2}\int^{\infty}_0 dt \left\langle {\bf v}^{\bot}_i(t)
\cdot {\bf v}^{\bot}_i(0) \right\rangle,\label{Dtr(per)}
\end{eqnarray}

\noindent where ${\bf v}^{\|}$ and ${\bf v}^{\bot}$ are the parallel and perpendicular components of the velocity ${\bf v}$ to the director $\hat{n}_d$, given by

\begin{equation}
{\bf v}^{\|}=({\bf v}\cdot\hat{n}_d)\hat{n}_d,  \hspace{0.5cm} \mbox{}  \hspace{0.5cm}
{\bf v}^{\bot}={\bf v}-{\bf v}^{\|}.
\label{v(par-per)}
\end{equation}

\noindent We have divided by 2 the expression (\ref{Dtr(per)}) to obtain and average of the two degrees of freedom of ${\bf v}^{\bot}$, the expression for $D_{tr}^{\|}$ has the contribution of only one degree of freedom. In the calculation of the diffusion coefficient we first obtained the parallel and perpendicular velocities, defined in Eqs. (\ref{v(par-per)}), then the integrals involved in Eqs. (\ref{Dtr}), (\ref{Dtr(par)}), and (\ref{Dtr(per)}) were evaluated by doing the summation over a correlation time of 300 $\Delta t^*$. 

\section{Main results}
In this section we present the results obtained in this work. Reduced units will be assumed hereafter. Pressure-density phase diagrams were obtained for Gay-Berne fluids along different isotherms as function of the parameter $\kappa'$. In order to validate the program we simulated the $(3,5,2,1)$ GB fluid studied by De Miguel et al. \cite{enrique1}. Results are presented in Fig. \ref{phase-3521} for temperatures $T^*=0.5$, 0.75, 1.00, and 1.25. We compared our results with those of Ref. \cite{enrique1} for the temperatures there reported and good agreement was found in all cases. In particular, the isotherm $T^*=0.5$, taken from \cite{enrique1}, is shown with stars in Fig. \ref{phase-3521} for comparison. At the lower temperature $T^*=0.5$, as the density increases the pressure increases for densities less than $\rho^*\sim 0.275$, then the pressure shows a decay for an intermediate region and eventually it increases again. For isotherms of higher $T^*$ a similar behavior can be observed, however the decay of pressure is shifted to regions of higher density and more than one decay can occur \cite{enrique1}. This effect is observed for $T^*=0.75$ and $T^*=1.0$ in the same figure. 

\begin{figure}[htp]
\centering\mbox{\includegraphics[width=4.0in]{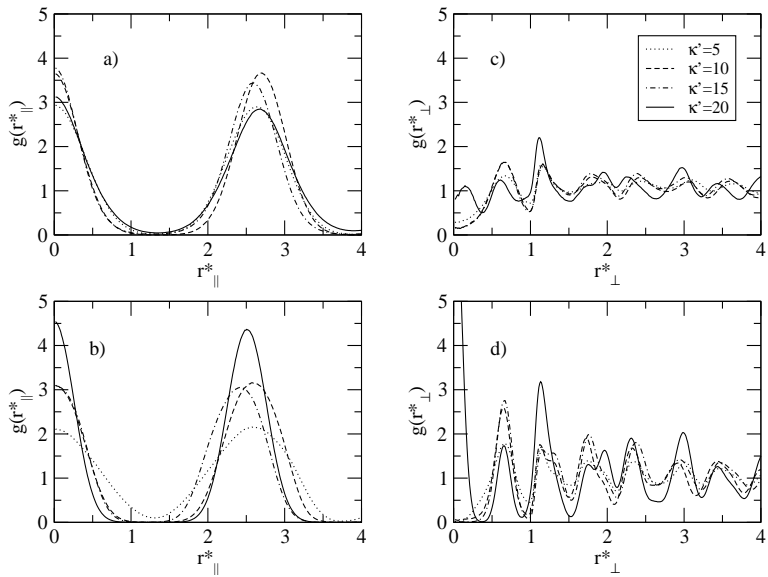}}
\caption{ Parallel pair correlation function $g(r^*_{\|})$ at temperature $T^*$ = 0.50 and densities: a) $\rho^*$ = 0.30 and b) $\rho^*$ = 0.35, for the PGB(3,$k'$,2,1) with $k'=5,10,15,20$. Perpendicular pair correlation function for the same conditions that parallel one at densities: c) $\rho^*$ = 0.30 and d) $\rho^*$ = 0.35.}
\label{gdr-3k21-T050}
\end{figure}

We explored the effect of changing $\kappa'$ in a systematic way, so the pressure was calculated for $\kappa'=5$, 10, 15, 20 for the temperatures studied in Fig. \ref{phase-3521}: $T^*=0.5$, 0.75, 1.00, and 1.25. Figure \ref{ppo-5101520} a)shows the pressure as function of density for $\kappa'=5$, 10, 15, 20 for the lower temperature  $T^*=0.5$. As we increase $\kappa'$ the region for the isotropic phase shifts to lower densities, meaning that a phase transition occurs before, for instance, for $\kappa'=20$ the decay of the pressure occurs at $\rho^* \sim 0.22$ while it decays around $\rho^* \sim 0.27$ for $\kappa'=5$, as seen from Fig. \ref{ppo-5101520} a). This effect is enhanced at low temperatures, as can be seen when we compared the pressure at temperature  $T^*=0.5$ in Fig. \ref{ppo-5101520} a) and that of higher temperature $T^*=1.0$ in \ref{ppo-5101520} b). For the same value of $\kappa'$ let say $\kappa'=5$, we observed that the decay on pressure takes place at $\rho^*\sim 0.29$ for $T^*=0.5$ while for $T^*=1.0$ this decay occurs at $\rho^* \sim 0.31$.

The order parameter $\langle P_2\rangle$ was evaluated for the isotherms already discussed and the values of $\kappa'=5$, 10, 15, 20. This quantity is shown in Figs. \ref{ppo-5101520} c) and \ref{ppo-5101520} d) at temperatures $T^*=0.5$ and $T^*$ = 1.00, respectively. In Fig. \ref{ppo-5101520} c) as $\kappa'$ increases the order parameter takes higher values for a fixed density. Along a given isotherm, $\langle P_2\rangle$ increases monotonically in the isotropic phase at low densities, then a sudden increase takes place for densities where the pressure decays, then $\langle P_2\rangle$ increases again and eventually the value of unity is reached. For the largest value of  $\kappa'$ this increase takes place at slightly lower densities. The differences found in the order parameter between states of equal density and different $\kappa'$ are more pronounced for low temperatures as can be observed when we compare results at $T^*$ = 0.5 (Fig. \ref{ppo-5101520} b)) and $T^*$ = 1.0 (Fig. \ref{ppo-5101520} c)). The differences in the order parameter as $\kappa'$ increases, are significantly reduced at temperature $T^*$ = 1.0 as is shown in Fig. \ref{ppo-5101520} c). 

\begin{figure}[htp]
\centering\mbox{\includegraphics[width=3.5in]{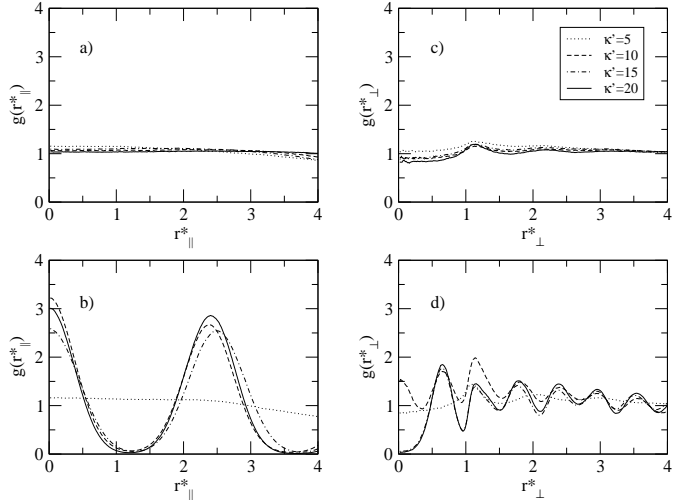}}
\caption{Same as Fig. \ref{gdr-3k21-T050} at a higher temperature: $T^*$ = 1.00 and densities a) $\rho^*$ = 0.30 and b) $\rho^*$ = 0.35 for the PGB(3,$k'$,2,1) with $k'=5,10,15,20$ as indicated on the inset for parallel correlation function. For perpendicular correlation function same conditions that parallel one at densities: c) $\rho^*$ = 0.30 and d) $\rho^*$ = 0.35.}
\label{gdr-3k21-T100}
\end{figure}

As an example of temperature effects, Fig. \ref{ppo-32021} a) shows the pressure-density curve for the (3,20,2,1) GB fluid at temperatures $T^*$ = 0.5, 0.75, 1.00, and 1.25. The corresponding order parameter is shown in Fig. \ref{ppo-32021} c), its behavior is consistent with the decays on the pressure shown in Fig. \ref{ppo-32021} a). A similar behavior as that for the $(3,5,2,1)$ GB was found in this $(3,20,2,1)$ GB fluid, however the region of densities for the isotropic phase shrinks and after the decay, the pressure takes lower values for $\kappa'=20$ than those of the $\kappa'=5$ case, as can be seen when we compare both cases in Figs. \ref{ppo-32021} b) at $T^*$ = 0.5 and $T^*$ = 1.0.

A comparison of the order parameter for the (3,5,2,1) and (3,20,2,1) GB fluids is shown in Fig. \ref{ppo-32021} d). The observed behavior confirms the findings showed in the pressure.

The parallel and perpendicular pair correlation functions, $g(r^*_{\|})$ and $g(r^*_{\bot})$, were obtained for different conditions of density, temperature and $\kappa'$. Figure \ref{gdr-3k21-T050} a) and b) show $g(r^*_{\|})$ at $\rho^*=0.3$ and $\rho^*=0.35$, respectively, at temperature $T^*$ = 0.5 for values $\kappa'=5, 10, 15, 20$ as indicated in the inset. The general trend is that the systems have already developed a layered structure at these conditions. At the lower density $\rho^*=0.3$ the $\kappa'=5$ and $\kappa'=20$ data for the pair correlation $g(r^*_{\|})$ did not show significant differences, something similar was found for the pressure as can be verified in Fig. \ref{ppo-32021} a). At this density the order parameter takes a slightly lower value for $\kappa'=5$ than for $\kappa'=20$. From this set the $\kappa'=10$ fluid showed the largest tendency to form the smectic phase at these conditions of density and temperature than the others, however the distance between layers is shorter in the $\kappa'=15$ fluid.

A different situation was found at $\rho^*=0.35$, Fig. \ref{gdr-3k21-T050} b), for $\kappa'=20$ the system shows a regular structure in $g(r^*_{\|})$, the maximums have the same height and the order parameter is closer to the value of unity,  however the inter-layer space is still shorter for the case $\kappa'=15$ or the structure is less defined. At this density the system with $\kappa'=5$ has less structure parallel to the director, which is opposite to the finding pointed out in Fig. \ref{gdr-3k21-T050} a).

\begin{figure}[htp]
\centering\mbox{\includegraphics[width=4.0in]{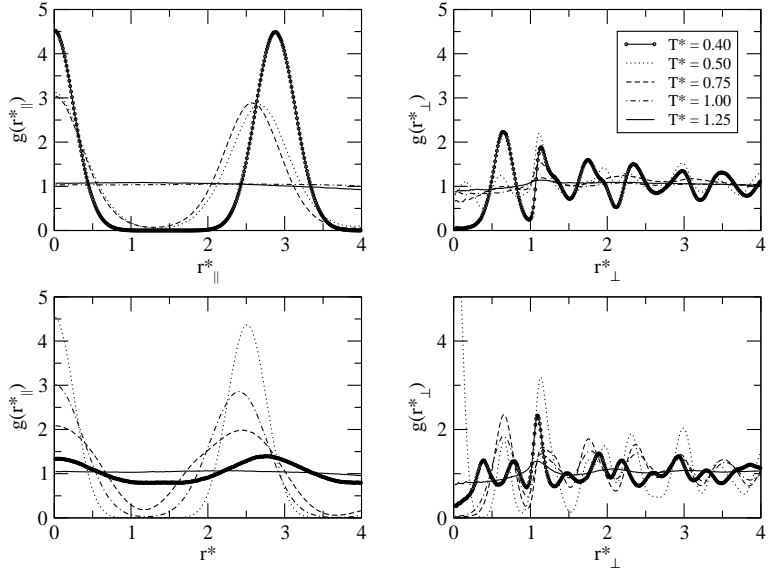}}
\caption{Parallel pair correlation function $g(r^*_{\|})$ at temperatures $T^*$ = 0.40, 0.50, 0.75, 1.00 and 1.25 and densities: a) $\rho^*$ = 0.30 and b) $\rho^*$ = 0.35, both for the PGB(3,20,2,1). Perpendicular pair correlation function $g(r^*_{\bot})$ at the same conditions that $g(r^*_{\|})$ at densities: c) $\rho^*$ = 0.30 and d) $\rho^*$ = 0.35}
\label{gdr-32021-T}
\end{figure}

The corresponding perpendicular correlation function for the systems presented in Figs. \ref{gdr-3k21-T050} a) and b) are shown in Figures \ref{gdr-3k21-T050} c) and d) at densities $\rho^*=0.3$ and $\rho^*=0.35$, respectively. This perpendicular correlations measure the intra-layer structure in the fluids. The structure inside a layer takes place at lower distances and have a longer range for $\kappa'=20$, than for $\kappa'=5, 10, 15$. At the higher density $\rho^*=0.35$ this effect is much more visible, the structure is well defined for  $\kappa'=20$ and a double shoulder at $r^*=1.8$ can be seen. In terms of the interaction, at this temperature $T^*=0.5$ the dominant configuration is the side-side as can be seen in Fig. \ref{pgb-01}, where the well depth is smaller for $\kappa'=20$ than for  $\kappa'=5$.

Figures \ref{gdr-3k21-T100} a) and b) shows $g(r^*_{\|})$ for the same systems as in Fig. \ref{gdr-3k21-T050} a) and b) but at a higher temperature $T^*=1.0$. At this temperature the layer structure is absent for all $\kappa'$ values studied as shown in Fig. \ref{gdr-3k21-T100} a) for $\rho^*=0.3$, just at this density the pressure loss the monotonic increase as can be observed in Fig. \ref{ppo-5101520} b) and the order parameter increases, see Fig. \ref{ppo-32021} b). For a higher density $\rho^*=0.35$ the layer structure takes place for $\kappa'=10, 15, 20$, while for $\kappa'=5$ is totally absent.

For the intra-layer structure we analyzed the perpendicular pair correlation $g(r^*_{\bot})$ for the same systems as in Fig. \ref{gdr-3k21-T050} c) and d) at temperature $T^*=1.0$, this is shown in Fig. \ref{gdr-3k21-T100} c) at $\rho^*=0.3$, for these conditions the structure is almost absent for all the $\kappa'$ values, while at $\rho^*=0.35$ the intra-layer is already well defined at shorter distances for $\kappa'=10$, for $\kappa'=15$ and $\kappa'=20$ it is about the same and for $\kappa'=5$ it is considerably reduced as shown in Fig. \ref{gdr-3k21-T100} d).

\begin{figure}[htp]
\centering\mbox{\includegraphics[width=3.50in]{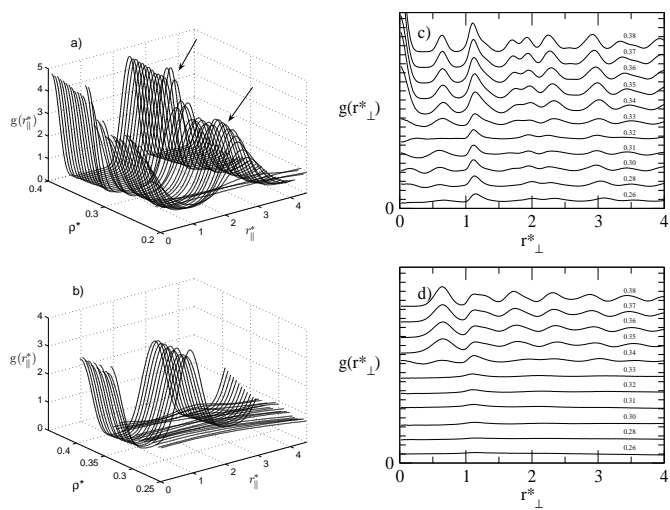}}
\caption{Parallel pair correlation function $g(r^*_{\|})$ at temperatures: a) $T^*$ = 0.50 and b) $T^*$ = 1.00 for densities between $\rho^*=0.26-0.38$ for the (3,20,2,1) GB fluid. Arrows on figure a) indicate regions where the function shows a decrease as an indicative of different transitions. Parallel pair correlation function $g(r^*_{\bot})$ at temperatures: c) $T^*$ = 0.50 and d) $T^*$ = 1.00 for the same conditions that the parallel one. The densities are indicate on the figure.}
\label{gdr-32021-RHO}
\end{figure}

In order to investigate the effect of temperature we considered isotherms at $T^*=0.4, 0.5, 0.75, 1.00, 1.25$ for the $(3,20,2,1)$ GB fluid. The parallel and perpendicular pair correlation functions are presented in Fig. \ref{gdr-32021-T} a) and c) respectively, as function of temperature for density $\rho^*=0.3$, and b) and d) for $\rho^*=0.35$. At $T^*=0.4$ and $\rho^*=0.3$ the layers can be well identified as can be seen from \ref{gdr-32021-T} a); as temperature increases the layer order decreases and eventually it disappears, for instance at $T^* \ge 1.0$ it is completely absent. Looking at the intra-layer structure with $g(r^*_{\bot})$ in Fig. \ref{gdr-32021-T} c) we observed a well developed layer structure which takes place at shorter distances for  $T^*=0.5$ than for $T^*=0.4$, but at this lower temperature the order by layer is quite considerable as compared to higher temperatures.
 
In Fig. \ref{gdr-32021-T} b) we observed that at density $\rho^*=0.35$ the order by layers manifests at temperatures $T^*=0.5, 0.75, 1.00$, although it is not well developed at $T^*=0.4$ and it disappears at $T^*=1.25$. The intra-layer order manifest itself for all the temperatures except at the lowest $T^*=0.4$, as seen in Fig. \ref{gdr-32021-T} d).

The parallel pair correlations at these conditions are shown in Fig. \ref{gdr-32021-RHO} at a) $T^*=0.5$ and b) $T^*=1.0$ for densities $\rho^*=0.26-0.38$. The arrows in Fig. \ref{gdr-32021-RHO} a) indicate regions where the function $g(r_{\|})$ shows an increase in its amplitude as an indicative of an smectic B phase in the system \cite{enrique1}. The perpendicular pair correlations are shown in Fig. \ref{gdr-32021-RHO} for c)  $T^*=0.5$ and d) $T^*=1.00$, for densities $\rho^*=0.26-0.38$ as indicated on each figure. As the density increases the intra-layer order increases being more notorious at $T^*=0.5$. 

\begin{figure}[htp]
\centering\mbox{\includegraphics[width=2.0in]{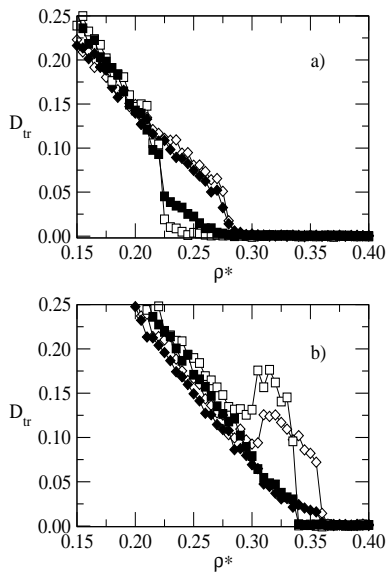}}
\caption{Parallel $D_{tr(\|)}$ (open symbols) and perpendicular $D_{tr(\bot)}$ (full symbols) diffusion coefficients as functions of density at two fixed temperatures: a) $T^*$ = 0.50 and b) $T^*$ = 1.00, for $\kappa'$ = 5 (diamonds) and $\kappa'=$ 20 (circles).}
\label{difusion}
\end{figure}

Finally, the parallel and perpendicular diffusion coefficients, $D_{tr\|}$ and $D_{tr\bot}$ as a function of density and temperature are presented as function of density for values $\kappa'=5, 20$ in Figs. \ref{difusion} a) at temperature $T^*=0.5$ and b) at $T^*=1.00$. As density increases both, $D_{tr\|}$ and $D_{tr\bot}$ decrease. However some particular features can be observed, for instance, for $\kappa'=20$ at densities $\rho^* < 0.22$ $D_{tr\|}$ and $D_{tr\bot}$ take the same value within the statistical error, but in the region $\rho^* = 0.22 - 0.28$ the diffusion is larger in the perpendicular than in the parallel direction. For $\kappa'=5$ the opposite behavior can be observed. At $\rho^* > 0.28$ the diffusion coefficient decays fast for both directions. At temperature $T^*=1$ a more complex behavior is found. Again there is a region where structural transitions take place and a non-systematic behavior in the diffusion is observed.

\section{Conclusions}

In this work we have studied Gay-Berne fluids by molecular dynamics simulations. Extensive simulations were performed to generate data for the pressure-density phase diagram, the orientational order parameter, the pair correlation functions and the translational diffusion coefficients via the velocity auto-correlation function. We studied Gay-Berne fluids with $\kappa=5$, $\mu=2$, $\nu=1$ and $\kappa'=5, 10, 15, 20$ at different conditions of density and temperature. The structure was analyzed in terms of the order parameter and the pair correlations, both parallel and perpendicular to the director. We explored the dependence of the thermodynamics and structural properties on changing the energy parameter $\kappa'$. Along a given isotherm, as density increases and the parameters ($\kappa$,$\kappa'$,$\mu$,$\nu$)  are kept constant the system is forced to the ordering undergoing transitions to different ordered phases.

Concerning the pressure when we fixed $T^*$, the general behavior is that pressure exhibits a monotonically increase in the low-density region, followed by several decays, the number of which depends on the $\kappa'$ value. The first decay is shifted to lower densities as $\kappa'$ increases. This effect was observed for all $\kappa'$ studied. This is in agreement with the behavior of the order parameter $\langle P_2 \rangle$ and the pair distribution functions, which was also confirmed by the diffusion coefficients, which are in particular shown in Fig. \ref{difusion} for $\kappa'=5$ and $20$. In addition, we explored the effect of changing the temperature when we fixed the GB parameters ($\kappa$,$\kappa'$,$\mu$,$\nu$) at constant $\rho^*$. We observed that the increase in temperature can suppress some of the structural phases, which in turns leads to the pressure to exhibit less decays as can be seen when we compared Fig. \ref{ppo-5101520} at a lower temperature with the results showed in Fig. \ref{ppo-5101520} b). 

From results on the parallel pair distribution function we believe that for high values of $\kappa'$ more than one smectic B phase can occur. In particular, when we simulated the GB fluid with $\kappa'=20$ the parallel pair correlation function showed increases in the amplitude of the maximums as density was increased. A non-monotonic behavior of $g(r_{\|})$ surrounding two regions of higher values is seen in the second maximum, as indicated with arrows in Fig. \ref{gdr-32021-RHO} a). 

Results on both parallel and perpendicular translational diffusion coefficients were obtained for a wide range of densities under different conditions of temperature and for different values of $\kappa'$. They can help in the description of the ordered phases.

We would like to mention that all the data here reported can be used to classify different structural phases present in Gay-Berne fluids. Additional work is needed to complete this task but nevertheless they can be used in combination with thermodynamic integration for these purposes.

\section*{Acknowledgements}

The authors gratefully acknowledge supercomputer facilities of Laboratorio Nacional de Superc\'omputo del Suroeste de M\'exico 
(LNS) proyect 201801014N1R y 201901004N.

\vspace{1.6in}

\newpage
\end{document}